\documentclass[a4paper,10pt]{article}
\usepackage{graphicx,amsmath,natbib}
\def\aap{A\&A\,  }
\def\apj{ApJ\,  }
\def\apjl{ApJ\,  }
\def\apjs{ApJS  }
\def\apss{Astrophysics and Space Science  }

\def\mnras{MNRAS\,  }

\def\sovast{Soviet Astronomy} 
\def\sn1987a{SN \,1987A\,}
\def\sun{\hbox{$\odot$}}
\DeclareRobustCommand{\orderof}{\ensuremath{\mathcal{O}}}
\def\snr{SN\,1993J~}
\def\degreezan{^{\,\circ}}
\def\aap{A\&A\,  }
\def\apj{ApJ\,  }
\def\apjl{ApJ\,  }
\def\apjs{ApJS  }
\def\apss{Astrophysics and Space Science  }

\def\mnras{MNRAS\,  }

\def\sovast{Soviet Astronomy} 
\title
{
A classical and a relativistic law   of
motion for  SN1987A
}
\author{L. Zaninetti   \\
Dipartimento di Fisica\\
Via Pietro Giuria 1
10125, Turin, Italy    \\
\footnote{zaninetti@ph.unito.it}\hspace{0.08cm}
{Corresponding author: zaninetti@ph.unito.it}}

\begin{document}
\maketitle
\begin{abstract}
In this paper we derive  some  first order differential
equations which model    the classical and the relativistic
thin layer approximations
in the presence of a circumstellar medium with a density
which is decreasing in the  distance $z$
from the equatorial
plane.
The circumstellar medium is
assumed to follow  a density profile with $z$ of
hyperbolic  type,
power law   type,
exponential type
or Gaussian type.
The first order differential equations
are solved analytically,
or numerically, or by a series  expansion,
or by
Pad\'e  approximants.
The initial conditions are chosen in order to model
the temporal evolution of SN 1987A   over 23  years.
The free parameters of the theory are found
by maximizing the observational reliability
which is based on an observed section of SN 1987A.
\end{abstract}

\section{Introduction}

The theories of the expansion of supernovae (SN) in the
circumstellar medium (CSM) are usually built in a spherical
framework.
Unfortunately, only a few SNs present a spherical expansion, such as
\snr , see \cite{Marcaide2009,MartiVidala2011}.
The more common observed morphologies are barrel or
hourglass shapes, see \cite{Lopez2014} for a classification.
A possible classification for the asymmetries firstly identifies
the center of the explosion and then defines the radius in the
equatorial plane, $R_{eq}$,
then the radius in the downward direction, $R_{down}$,
and then the radius in the upward direction, $R_{up}$,
see \cite{Zaninetti2000}.
The above classification allows introducing  a symmetry:
$R_{down}$=$R_{up}$ means that the expansion from
the equatorial plane along the two opposite
polar directions is the same.
A second symmetry can be introduced in
the framework of spherical coordinates
assuming independence from the azimuthal angle.
The theories for asymmetric SNs or late
supernova remnants (SNRs) have therefore been set up,
we select some of them.
Possible reasons for the distortion of SNRs
have been extensively
studied analytically and numerically by
\cite{Chevalier1974,Bisnovatyi-Kogan1989,Igumenshchev1992,Arthur1993,Maciejewski1999}.
Two SNRs presenting a barrel morphology were observed  and
explained in \cite{Gaensler1998}.
Numerical calculations of the interaction of an SN with an
axisymmetric structure with a high density
in the equatorial plane  were carried out by
\cite{Blondin1996}.

New  laws  of  motion,
assuming  that  only  a  fraction  of  the  mass  which  resides in the surrounding medium is accumulated in the
advancing thin layer, were developed
both in a classical framework in the presence of  an exponential profile,
see \cite{Zaninetti2012b} or  an isothermal self-gravitating disk,
see \cite{Zaninetti2013c},
and both in a relativistic framework in the presence of
an isothermal self-gravitating disk,
see \cite{Zaninetti2014c}.
We now present some maximum observed velocities in SNs:
the maximum velocity  for  Si II $\lambda$6355
vary in [15000,25000]\ km s$^{-1}$  according to Figure 13 in
\cite{Silverman2015}
or in  [13000,24000]\ km s$^{-1}$
according to Figure 4 in \cite{Zhao2015}.
These high observed velocities  demand a relativistic
treatment of the theory.
In this paper we introduce, in
Section \ref{secpreliminary},
four  asymmetric density profiles,
in  Section \ref{secclassical}
we derive the  differential equations
which model the thin layer approximation
for an  SN in the presence of four asymmetric types   of medium
and a   relativistic treatment is carried out in
Section \ref{secrelativistic}
for two asymmetric types   of medium.

\section{Preliminaries}
\label{secpreliminary}

This  section  introduces
the spherical coordinates and
four  density profiles with axial symmetry
for the CSM:
a hyperbolic  profile,
a power law profile,
an exponential profile
and a Gaussian profile.

\subsection{Spherical coordinates}

\begin{figure*}
\begin{center}
\includegraphics[width=7cm]{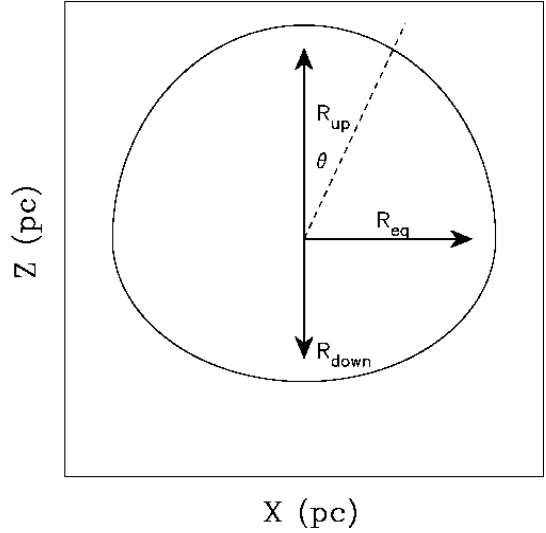}
\end {center}
\caption
{
Sketch for an asymmetric SN.
}
\label{section_asymm}
    \end{figure*}

A point in  Cartesian coordinates is characterized by
$x,y$, and $z$.
The same point in spherical coordinates is characterized by
the radial distance $r \in[0,\infty]$,
the polar angle  $\theta \in [0,\pi]$,
and the azimuthal angle $\varphi \in [0,2\pi]$.
Figure \ref{section_asymm} presents a section  of an
asymmetric SN where there can be clearly seen
the polar  angle
$\theta$ and the three
observable radii $R_{up}$, $R_{down}$, and $R_{eq}$.

\subsection{A hyperbolic  profile}

The density  is  assumed to have the following
dependence on $z$
in Cartesian coordinates,
\begin{equation}
 \rho(z;z_0,\rho_0) =
  \begin{cases}
    \rho_0                 & \quad \text{if } z   \leq z_0\\
    \rho_0 \frac{z_0}{z}   & \quad \text{if } z   >    z_0\\
  \end{cases}
  \label{profhyperbolic}
\quad ,
\end{equation}
where the parameter $z_0$ fixes the scale and  $\rho_0$ is the
density at $z=z_0$.
In spherical coordinates
the dependence  on the polar angle is
\begin{equation}
 \rho(r;\theta,z_0,\rho_0) =
  \begin{cases}
    \rho_0                 & \quad \text{if } r \cos(\theta)  \leq z_0\\
    \rho_0 \frac{z_0}{r \cos(\theta)}   & \quad \text{if } r \cos(\theta)  >    z_0\\
  \end{cases}
  \label{profhyperbolicr}
  \quad  .
\end{equation}

Given a solid angle  $\Delta \Omega$
the mass $M_0$ swept
in the interval $[0,r_0]$
is
\begin{equation}
M_0 =
\frac{4}{3}\,\rho_{{0}}\pi \,{r_{{0}}}^{3} \Delta \Omega
\quad .
\end{equation}
The total mass swept, $M(r;r_0,z_0,\alpha,\theta,\rho_0) $,
in the interval $[0,r]$
is
\begin{equation}
M(r;r_0,z_0,\alpha,\theta,\rho_0)= \bigl (
\frac{4}{3}\,\rho_{{0}}\pi \,{r_{{0}}}^{3}+2\,{\frac {\rho_{{0}}{\it z_0}\,\pi
\, \left( {r}^{2}-{r_{{0}}}^{2} \right) }{\cos \left( \theta \right) }
}
\bigr) \Delta \Omega
\quad .
\label{masshyperbolic}
\end{equation}
The density $\rho_0$ can be  obtained
by introducing  the number density  expressed  in particles
$\mathrm{cm}^{-3}$,
$n_0$,
the mass of  hydrogen, $m_H$,
and  a multiplicative factor $f$,
which is chosen to be  1.4, see \cite{Dalgarno1987},
\begin{equation}
\rho_0  = f  m_H n_0
\quad .
\end{equation}
The astrophysical version of the total swept mass,
expressed in solar mass
units, $M_{\sun}$,  is therefore
\begin{equation}
M (r_{pc};z_{0,pc},n_0,\theta)\approx
\frac{0.216 \,n_{{0}}{\it z}_{{{\it 0,pc}}}{r_{{{\it pc}}}}^{2}}
{cos(\theta)}
\Delta \Omega \,M_{\sun}
\quad ,
\end{equation}
where
$z_{0,pc}$, $r_{0,pc}$ and $r_{0,pc}$
are  $z_0$,      $r_0$      and $r$ expressed  in pc units.

\subsection{A power law profile}

The density  is  assumed to have the following dependence on $z$
in Cartesian coordinates:
\begin{equation}
 \rho(z;z_0,\rho_0) =
  \begin{cases}
    \rho_0                            & \quad \text{if } z   \leq z_0\\
    \rho_0 (\frac{z_0}{z})^{\alpha}   & \quad \text{if } z   >    z_0\\
  \end{cases}
  \label{profpower}
\quad ,
\end{equation}
where $z_0$ fixes the scale.
In spherical coordinates,
the dependence  on the polar angle is
\begin{equation}
 \rho(r,\theta,z_0,\rho_0) =
  \begin{cases}
    \rho_0                                         & \quad \text{if } r \cos(\theta)  \leq z_0\\
    \rho_0 (\frac{z_0}{r \cos(\theta)})^{\alpha}   & \quad \text{if } r \cos(\theta)  >    z_0\\
  \end{cases}
  \label{profpowerz}
\end{equation}
The mass $M_0$ swept
in the interval  $[0,r_0]$ in a given solid angle
is
\begin{equation}
M_0 =
\frac{4}{3}\,\rho_{{0}}\pi \,{r_{{0}}}^{3} \Delta \Omega
\quad .
\end{equation}
The total mass swept,
$M(r;r_0,\alpha,\theta,\rho_0) $,
in the interval $[0,r]$
is
\begin{multline}
M (r;r_0,\alpha,z_0,\theta,\rho_0)=  \\
\bigg (
\frac{4}{3}\,\rho_{{0}}\pi \,{r_{{0}}}^{3}-4\,{\frac {{r}^{3}\rho_{{0}}\pi }{
\alpha-3} \left( {\frac {{\it z_0}}{r\cos \left( \theta \right) }}
 \right) ^{\alpha}}+4\,{\frac {\rho_{{0}}\pi \,{r_{{0}}}^{3}}{\alpha-3
} \left( {\frac {{\it z_0}}{r_{{0}}\cos \left( \theta \right) }}
 \right) ^{\alpha}}
\bigg ) \Delta \Omega
\quad .
\label{sweptmasspower}
\end{multline}
The astrophysical  swept mass is
\begin{equation}
M (r_{pc};z_{0,pc},\alpha,n_0,\theta )\approx
\frac
{
0.432 \,n_{{0}}{{\it z}_{{{\it 0,pc}}}}^{\alpha}{r_{{{\it pc}
}}}^{-\alpha+3} \left( \cos \left( \theta \right)  \right) ^{-\alpha}
}
{
3-\alpha
}
\Delta \Omega \,M_{\sun}
\quad .
\end{equation}

\subsection{An exponential profile}

The density  is  assumed to have the following
exponential dependence on $z$
in Cartesian coordinates:
\begin{equation}
 \rho(z;b,\rho_0) =
\rho_0  \exp{(-z/b)}
\quad ,
\label{profexponential}
\end{equation}
where $b$ represents the scale.
In spherical coordinates,
the density is
\begin{equation}
 \rho(r;r_0,b,\rho_0) =
  \begin{cases}
    \rho_0                                 & \quad \text{if } r    \leq r_0\\
    \rho_0 \exp{-\frac{r\cos(\theta)}{b}}  & \quad \text{if } r    >   r_0\\
  \end{cases}
  \label{profexponentialr}
\end{equation}
The total mass swept,   $M(r;r_0,b,\theta,\rho_0) $,
in the interval $[0,r]$ is
\begin{multline}
M (r;r_0,b,\theta,\rho_0)= \\
\frac{4}{3}\,\rho_{{0}}\pi \,{r_{{0}}}^{3}-4\,{\frac {b \left( {r}^{2} \left(
\cos \left( \theta \right)  \right) ^{2}+2\,rb\cos \left( \theta
 \right) +2\,{b}^{2} \right) \rho_{{0}}\pi }{ \left( \cos \left(
\theta \right)  \right) ^{3}}{{\rm e}^{-{\frac {r\cos \left( \theta
 \right) }{b}}}}}
\\
+4\,{\frac {b \left( {r_{{0}}}^{2} \left( \cos
 \left( \theta \right)  \right) ^{2}+2\,r_{{0}}b\cos \left( \theta
 \right) +2\,{b}^{2} \right) \rho_{{0}}\pi }{ \left( \cos \left(
\theta \right)  \right) ^{3}}{{\rm e}^{-{\frac {r_{{0}}\cos \left(
\theta \right) }{b}}}}}
 \Delta \Omega
\quad .
\label{massexponential}
\end{multline}
The astrophysical version expressed in solar masses is
\begin{multline}
M (r;r_{0,pc},b_{pc},\theta,n_0)=  \\
-\frac{1}{\left( \cos \left( \theta \right)  \right) ^{3}}
\biggl(
0.288\,n_{{0}}   ( - 0.5\,{{\it r}_{{{\it 0,pc}}}}^{3}
   ( \cos   ( \theta   )    ) ^{3} 
   + 1.5\,{\it b_{pc}}\,{
{\rm e}^{- 1\,{\frac {r_{{{\it pc}}}\cos   ( \theta   ) }{{
\it b_{pc}}}}}}{r_{{{\it pc}}}}^{2}   ( \cos   ( \theta   )
   ) ^{2}
\\
+ 3.0\,{{\it b_{pc}}}^{2}{{\rm e}^{- 1\,{\frac {r_{{{\it
pc}}}\cos   ( \theta   ) }{{\it b_{pc}}}}}}r_{{{\it pc}}}\cos
   ( \theta   )
+ 3.0\,{{\it b_{pc}}}^{3}{{\rm e}^{- 1\,{\frac {
r_{{{\it pc}}}\cos   ( \theta   ) }{{\it b_{pc}}}}}}
\\
- 1.5\,{\it
b_{pc}}\,{{\rm e}^{- 1\,{\frac {{\it r}_{{{\it 0,pc}}}\cos   (
\theta   ) }{{\it b_{pc}}}}}}{{\it r}_{{{\it 0,pc}}}}^{2}   ( \cos
   ( \theta   )    ) ^{2}
 \\
- 3.0\,{{\it b_{pc}}}^{2}{{\rm e}^{-
 1\,{\frac {{\it r}_{{{\it 0,pc}}}\cos   ( \theta   ) }{{\it
b_{pc}}}}}}{\it r}_{{{\it 0,pc}}}\cos   ( \theta   ) - 3.0\,{{\it
b_{pc}}}^{3}{{\rm e}^{- 1\,{\frac {{\it r}_{{{\it 0,pc}}}\cos   (
\theta   ) }{{\it b_{pc}}}}}}   )
\biggr )
\Delta \Omega \,M_{\sun}
\quad ,
\end{multline}
where $b_{pc}$ is the scale expressed in pc.

\subsection{A  Gaussian profile}

The density  is  assumed to have the following Gaussian
dependence on $z$
in Cartesian coordinates:
\begin{equation}
 \rho(z;b,\rho_0) =
\rho_0  {{\rm e}^{-\frac{1}{2}\,{\frac {{z}^{2}}{{b}^{2}}}}}
\quad ,
\label{profgaussian}
\end{equation}
where $b$ represents the standard deviation.
In spherical coordinates,
the density is
\begin{equation}
 \rho(r;r_0,b,\rho_0) =
  \begin{cases}
    \rho_0                                               & \quad \text{if } r    \leq r_0\\
    \rho_0 {{\rm e}^{-\frac{1}{2}\,{\frac {{z}^{2}}{{b}^{2}}}}}  & \quad \text{if } r    >   r_0\\
  \end{cases}
  \label{profgaussianr}
\quad .
\end{equation}

The total mass swept,   $M(r;r_0,b,\theta,\rho_0) $,
in the interval $[0,r]$
is
\begin{multline}
M (r;r_0,b,\theta,\rho_0)=
\frac{4}{3}\,\rho0\,\pi \,{{\it r_0}}^{3}+4\,\rho0\,\pi \, \Bigl  ( -{\frac {r{b}
^{2}}{   ( \cos   ( \theta   )    ) ^{2}}{{\rm e}^{-\frac{1}{2}
\,{\frac {{r}^{2}   ( \cos   ( \theta   )    ) ^{2}}{{b}
^{2}}}}}}\\
+\frac{1}{2}\,{\frac {{b}^{3}\sqrt {\pi }\sqrt {2}}{   ( \cos
   ( \theta   )    ) ^{3}}{\rm erf}   (\frac{1}{2}\,{\frac {
\sqrt {2}\cos   ( \theta   ) r}{b}}  )} \Bigr  ) -4\,\rho0\,
\pi \, \Bigl  ( -{\frac {{\it r_0}\,{b}^{2}}{   ( \cos   ( \theta
   )    ) ^{2}}{{\rm e}^{-\frac{1}{2}\,{\frac {{{\it r_0}}^{2}   (
\cos   ( \theta   )    ) ^{2}}{{b}^{2}}}}}}\\
+\frac{1}{2}\,{\frac {{b
}^{3}\sqrt {\pi }\sqrt {2}}{   ( \cos   ( \theta   )
   ) ^{3}}{\rm erf}   (\frac{1}{2}\,{\frac {\sqrt {2}\cos   ( \theta
   ) {\it r_0}}{b}}  )} \Bigr  )
 \Delta \Omega
\quad ,
\label{massgaussian}
\end{multline}

where $\mathop{\mathrm{erf}}(x)$
is the error function, defined by
\begin{equation}
\mathop{\mathrm{erf}\/}\nolimits
(x)=\frac{2}{\sqrt{\pi}}\int_{0}^{x}e^{-t^{2}}dt
\quad .
\end{equation}

The previous formula expressed is solar masses is
\begin{multline}
M (r;r_{0,pc},b_{pc},\theta,n_0)= \\
\frac{1}{    ( \cos    ( \theta    )     ) ^{3}}
\Biggl (
- 1.024\,10^{-7}\,n_{{0}}  \Bigl  ( - 1.4\,10^6\,{{\it r}_{{{\it 0,pc}}}}
^{3}    ( \cos    ( \theta    )     ) ^{3}\\
+ 4.226\,10^6\,{
{\rm e}^{- 0.5\,{\frac {{r_{{{\it pc}}}}^{2}    ( \cos    (
\theta    )     ) ^{2}}{{{\it b_{pc}}}^{2}}}}}r_{{{\it pc}}}{{\it
b_{pc}}}^{2}\cos    ( \theta    ) \\
- 5.297\,10^6\,{{\it b_{pc}}}^{3}
{\rm erf}    ( 0.707 \,{\frac {\cos    ( \theta    ) r_{{
{\it pcc}}}}{{\it b_{pc}}}}   )\\
- 4.226 \,10^6 \,{{\rm e}^{- 0.5\,{\frac {{
{\it r}_{{{\it 0,pc}}}}^{2}    ( \cos    ( \theta    )
    ) ^{2}}{{{\it b_{pc}}}^{2}}}}}{\it r}_{{{\it 0,pc}}}{{\it b_{pc}}}^{2
}\cos    ( \theta    ) \\
 + 5.297\,10^6\,{{\it b_{pc}}}^{3}{\rm erf}
   ( 0.707\,{\frac {\cos    ( \theta    ) {\it r}_{{{
\it pc}}}}{{\it b_{pc}}}}   )   \Bigr  )
\Biggr )
\Delta \Omega \,M_{\sun}
\quad .
\end{multline}

\section{The classical thin layer approximation}
\label{secclassical}

The conservation of the classical momentum in
spherical coordinates
along  the  solid angle  $\Delta \Omega$
in the framework of the thin
layer approximation  states that
\begin{equation}
M_0(r_0) \,v_0 = M(r)\,v
\quad ,
\end{equation}
where $M_0(r_0)$ and $M(r)$ are the swept masses at $r_0$ and $r$,
and $v_0$ and $v$ are the velocities of the thin layer at $r_0$ and $r$.
This conservation law can be expressed as a differential equation
of the first order by inserting $v=\frac{dr}{dt}$:
\begin{equation}
M(r)\, \frac{dr}{dt} - M_0\, v_0=0
\quad .
\end{equation}
The above differential equation is independent
of the azimuthal angle $\varphi$.
The 3D surface which represents the advancing shock of the SN
consists of the rotation  about the $z$-axis
of the curve in the $x-z$ plane defined by
the analytical or numerical solution $r(t)$;
this is the {\it first symmetry}.
A  {\it second symmetry} around the $z=0$ plane
allows building the two lobes of the advancing surface.
The orientation  of the 3D surface  is characterized by
the
Euler angles $(\Theta, \Phi, \Psi)$
and  therefore  by a total
$3 \times 3$  rotation matrix,
$E$, see \cite{Goldstein2002}.

The adopted astrophysical units  are pc
for length and
yr for time;
the initial velocity $v_{{0}}$ is
expressed in
pc yr$^{-1}$.
The astronomical velocities are evaluated in
\ km s$^{-1}$
and  therefore
$v_{{0}} =1.02\times10^{-6} v_{{1}}$
where  $v_{{1}}$ is the initial
velocity expressed in
km s$^{-1}$.

\subsection{The case of \sn1987a}

\begin{figure*}
\begin{center}
\includegraphics[width=7cm]{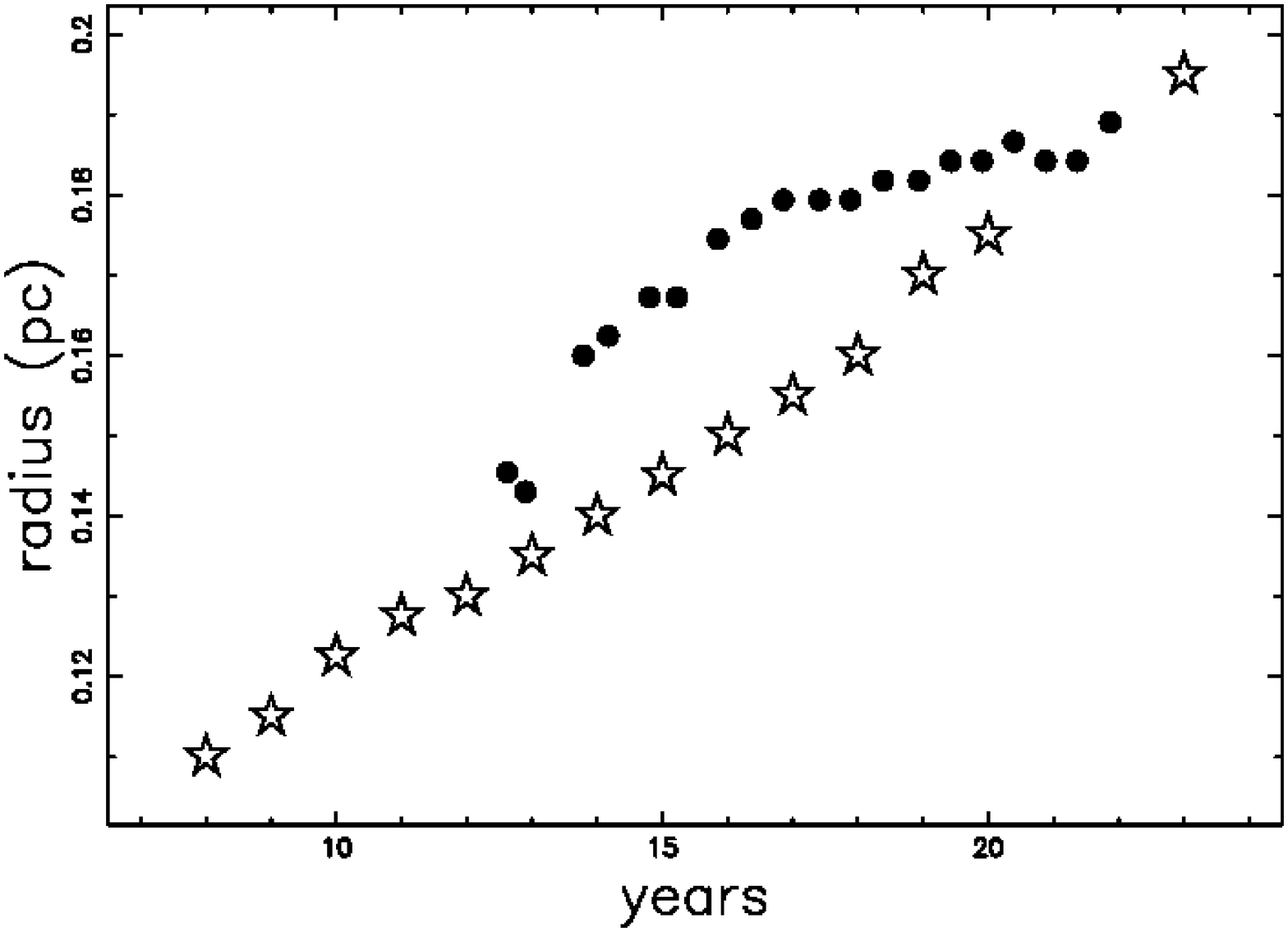}
\end {center}
\caption
{
Observed  radius of torus only  as function of time:
  full points as  in Racusin et al. 2009
  and
  empty stars as  in Chiad et al.  2012.
}
\label{twodataeq}
    \end{figure*}

\begin{figure*}
\begin{center}
\includegraphics[width=7cm]{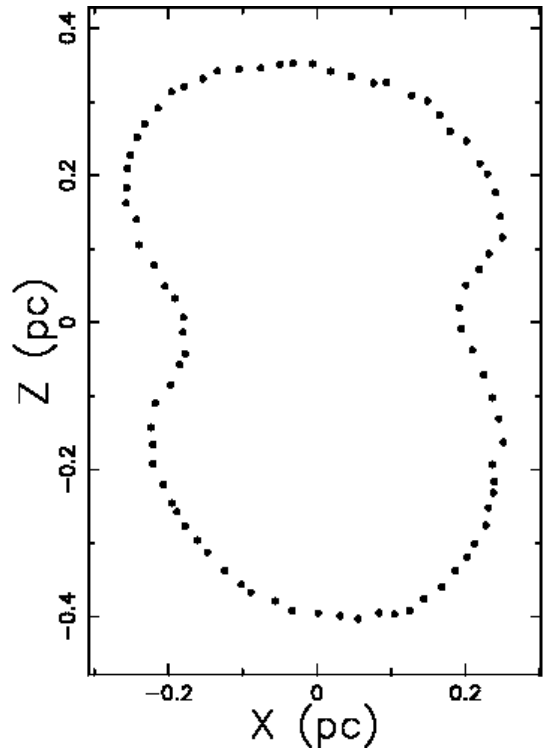}
\end {center}
\caption
{
Section of \sn1987a
in the $x-z$ plane adapted by the author from Figure 5
in \cite{France2015}.
}
\label{section_obs_sn1987a}
    \end{figure*}

The complex structure of \sn1987a can be classified as
a torus only, a torus plus two lobes, and a torus plus 4 lobes,
see \cite{Racusin2009}.
The region connected with the
radius of the advancing torus is here identified
with  our equatorial
region,
in  spherical coordinates, $\theta =\frac{\pi}{2}$.
The radius of the torus only  as a function of time can be found in
Table 2 of \cite{Racusin2009} or Figure 3 of \cite{Chiad2012},
see Figure  \ref{twodataeq} for a comparison of the
two different techniques.
The radius of the torus only as given by the
counting pixels method (\cite{Chiad2012})
shows a more regular behavior
and we calibrate our codes in the equatorial region
in such a way that at time
23 years   the radius is $\frac{0.39}{2}$\,pc.
Another useful resource for calibration is a section
of \sn1987a
reported as a sketch in Figure 5 of \cite{France2015}.
This section
was digitized  and  rotated in the $x-z$ plane
by $-40\degreezan$,  see Figure \ref{section_obs_sn1987a}.
The above approximate section allows introducing
an observational
percentage reliability, $\epsilon_{\mathrm {obs}}$,
over the whole range
of the polar   angle  $\theta$,
\begin{equation}
\epsilon_{\mathrm {obs}}  =100(1-\frac{\sum_j |r_{\mathrm {obs}}-r_{\mathrm{num}}|_j}{\sum_j
{r_{\mathrm {obs}}}_{,j}})
,
\label{efficiencymany}
\end{equation}
where
$r_{\mathrm{num}}$ is the theoretical radius,
$r_{\mathrm{obs}}$ is the observed    radius, and
the  index $j$  varies  from 1 to the number of
available observations, in our case 81.
The above statistical method allows fixing the parameters
of the theory in a scientific way rather
than adopting an ``ad hoc'' hypothesis.

\subsection{Motion with an hyperbolic profile}

\begin{figure*}
\begin{center}
\includegraphics[width=7cm]{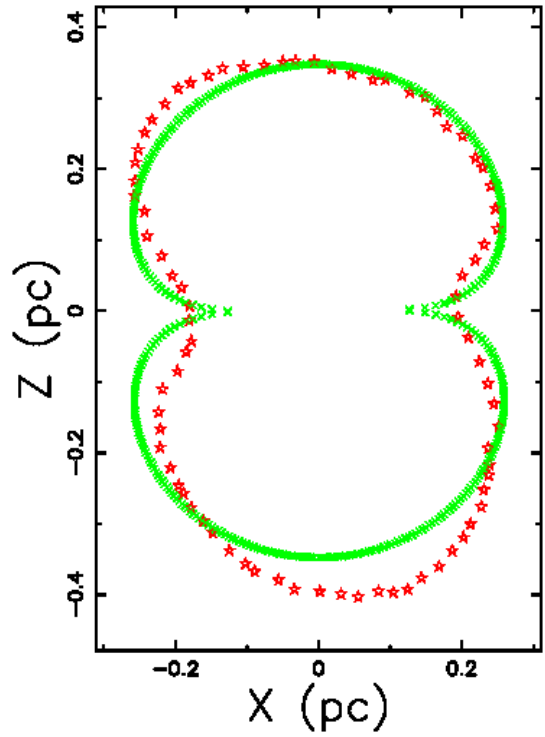}
\end {center}
\caption
{
Section of \sn1987a
in the $x-z$ plane with a hyperbolic profile
(green points)
and observed profile (red stars).
The parameters
$r_0=0.06$\ pc,
$z_0=0.001$\ pc,
$t=21.86$\ yr,
$t_0 =0.1$ yr  and
$v_0\,=25000$ km s$^{-1}$
give
$\epsilon_{\mathrm {obs}}=92.13\%$.
}
\label{cut_hyper_1987a}
    \end{figure*}

\begin{figure*}
\begin{center}
\includegraphics[width=7cm]{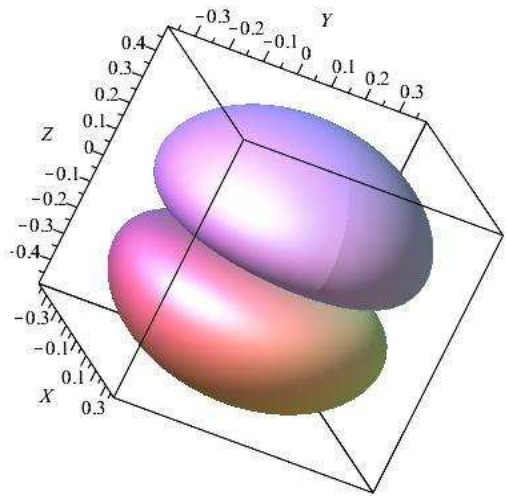}
\end {center}
\caption
{
3D surface  of  \sn1987a
with parameters as in Figure \ref{cut_hyper_1987a}.
The three Euler angles are $\Theta=40$, $\Phi=60$ and
$ \Psi=60 $.
}
\label{3dsurfacehyper}
    \end{figure*}

In the case of a hyperbolic density profile
for the CSM
as given by Eq. (\ref{profhyperbolic}),
the differential equation
which models momentum conservation
is
\begin{equation}
 \left(\frac{4}{3}\,\rho_{{0}}\pi \,{r_{{0}}}^{3}+2\,{\frac {\rho_{{0}}{\it
z_0}\,\pi \, \left( -{r_{{0}}}^{2}+ \left( r \left( t \right)  \right)
^{2} \right) }{\cos \left( \theta \right) }} \right) {\frac {\rm d}{
{\rm d}t}}r \left( t \right) -\frac{4}{3} \,\rho_{{0}}\pi \,{r_{{0}}}^{3}v_{{0}
}=0
\quad ,
\end{equation}
where the initial  conditions
are  $r=r_0$  and   $v=v_0$
when $t=t_0$.
The variables can be separated and
the radius as a function of the time
is
\begin{equation}
r(t;t_0,z_0,v_0)= \frac{HN}{HD}
\quad ,
\nonumber
\label{rtanalyticalhyper}
\end{equation}
where
\begin{multline}
HN= 
r_{{0}}\sqrt [3]{3}    ( -2\,\cos    ( \theta    ) \sqrt [3]{3}
r_{{0}}+    ( -9\,\cos    ( \theta    ) t_{{0}}v_{{0}}\sqrt {{
\it z_0}}+9\,\cos    ( \theta    ) tv_{{0}}\sqrt {{\it z_0}}
\\
+9\,
\cos    ( \theta    ) r_{{0}}\sqrt {{\it z_0}}
\\
-9\,{{\it z_0}}^{3/2}
+\sqrt {3}\sqrt {\cos    ( \theta    ) {\it AHN}}    ) ^{2/3}+
3\,\sqrt [3]{3}{\it z_0}    )
\end{multline}
with
\begin{multline}
AHN= 
27\,\cos    ( \theta     ) {t}^{2}{v_{{0}}}^{2}{\it z_0}-54\,\cos
    ( \theta     ) tt_{{0}}{v_{{0}}}^{2}{\it z_0}+27\,\cos    (
\theta     ) {t_{{0}}}^{2}{v_{{0}}}^{2}{\it z_0}
\\
+8\,    ( \cos
    ( \theta     )      ) ^{2}{r_{{0}}}^{3}
+54\,\cos    (
\theta     ) r_{{0}}tv_{{0}}{\it z_0}
\\
-54\,\cos    ( \theta     )
r_{{0}}t_{{0}}v_{{0}}{\it z_0}-9\,\cos    ( \theta     ) {r_{{0}}}^
{2}{\it z_0}
-54\,tv_{{0}}{{\it z_0}}^{2}+54\,t_{{0}}v_{{0}}{{\it z_0}}^{2
}
\end{multline}
and
\begin{equation}
HD =
3\,\sqrt {z_{{0}}}\sqrt [3]{{\it BHD}}
\end{equation}
with
\begin{multline}
BHD=
-9\,\cos \left( \theta \right) t_{{0}}v_{{0}}\sqrt {z_{{0}}}+9\,\cos
 \left( \theta \right) tv_{{0}}\sqrt {z_{{0}}}+9\,\cos \left( \theta
 \right) r_{{0}}\sqrt {z_{{0}}}
\\
-9\,{z_{{0}}}^{3/2}+\sqrt {3}\sqrt {
\cos \left( \theta \right) {\it AHN}}
\quad .
\end{multline}
The velocity  as a function of the radius $r$  is
\begin{equation}
v(t)=
2\,{\frac {{r_{{0}}}^{3}v_{{0}}\cos \left( \theta \right) }{2\,{r_{{0}
}}^{3}\cos \left( \theta \right) -3\,{r_{{0}}}^{2}z_{{0}}+3\,{r}^{2}z_
{{0}}}}
\quad .
\end{equation}
Figure \ref{cut_hyper_1987a} displays a cut  of  \sn1987a
in the $x-z$ plane.
A rotation around  the
$z$-axis  of the previous  section allows
building a 3D surface, see
Figure \ref{3dsurfacehyper}.

\subsection{Motion with a power law profile}

\begin{figure*}
\begin{center}
\includegraphics[width=7cm]{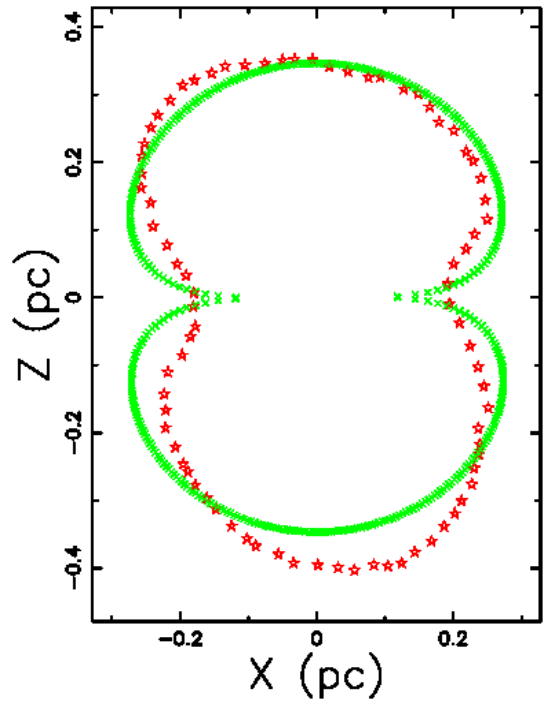}
\end {center}
\caption
{
Section of \sn1987a
in the $x-z$ plane with a power law profile
(green points)
and observed profile (red stars).
The parameters
$r_0=0.05$ pc,
$z_0=0.002$\ pc,
$t=21.86$\ yr,
$t_0 =0.1$\ yr,
$\alpha=1.3$,
$v_0\,=15000$ \ km s$^{-1}$
give
$\epsilon_{\mathrm {obs}}=90.76\%$.
}
\label{cut_power_1987a}
    \end{figure*}

In the case of a power-law density profile for the CSM
as given by Eq. (\ref{profpower}),
the differential equation
which models the momentum conservation
is
\begin{multline}
\frac{4}{3}\,\rho_{{0}}\pi \,{r_{{0}}}^{3}v_{{0}}
-\bigl (\frac{4}{3}\,\rho_{{0}}\pi
\,{r_{{0}}}^{3}-4\,{\frac {     ( r     ( t    )     ) ^{3}
\rho_{{0}}\pi }{\alpha-3}     ( {\frac {z_{{0}}}{r     ( t    )
\cos     ( \theta    ) }}    ) ^{\alpha}}\\
+4\,{\frac {\rho_{{0}}
\pi \,{r_{{0}}}^{3}}{\alpha-3}     ( {\frac {z_{{0}}}{r_{{0}}\cos
     ( \theta    ) }}    ) ^{\alpha}}  \bigr  ) {\frac {\rm d}{
{\rm d}t}}r     ( t    ) =0
\label{eqndiffpower}
\quad .
\end{multline}
The velocity is
\begin{equation}
v(r;r_0,v_0,\theta,\alpha) =
\frac
{
-{r_{{0}}}^{3}v_{{0}} \left( \alpha-3 \right)
}
{
3\, \left( {\frac {z_{{0}}}{r\cos \left( \theta \right) }} \right) ^{
\alpha}{r}^{3}-3\, \left( {\frac {z_{{0}}}{r_{{0}}\cos \left( \theta
 \right) }} \right) ^{\alpha}{r_{{0}}}^{3}-{r_{{0}}}^{3}\alpha+3\,{r_{
{0}}}^{3}
}
\quad .
\end{equation}
We now evaluate the following integral
\begin{equation}
I =\int_{r_0}^r \frac{1}{v(r;r_0,v_0,\theta,\alpha)} dr
\quad ,
\end{equation}
which is
\begin{multline}
I_p(r) =3\,{\frac {{r}^{4}}{{r_{{0}}}^{3}v_{{0}} \left( \alpha-3 \right)
 \left( \alpha-4 \right) }{{\rm e}^{\alpha\,\ln  \left( {\frac {z_{{0}
}}{r\cos \left( \theta \right) }} \right) }}}
\\
+3\,{\frac {r}{v_{{0}}
 \left( \alpha-3 \right) } \left( {\frac {z_{{0}}}{r_{{0}}\cos \left(
\theta \right) }} \right) ^{\alpha}}+{\frac {\alpha\,r}{v_{{0}}
 \left( \alpha-3 \right) }}-3\,{\frac {r}{v_{{0}} \left( \alpha-3
 \right) }}
\end{multline}
The solution of the differential equation (\ref{eqndiffpower})
can  be found solving numerically  the following nonlinear equation
\begin{equation}
I(r)- I(r_0)= t-t_0
\quad  .
\end{equation}
More precisely we used the
FORTRAN SUBROUTINE
ZBRENT from \cite{press} and
Figure \ref{cut_power_1987a} reports  the numerical
solution as a  cut  of  \sn1987a
in the $x-z$ plane.

\subsection{Motion with an exponential profile}

In the case of  an exponential density profile
for the CSM
as given by Eq. (\ref{profexponential}),
the differential equation
which models momentum conservation
is
\begin{multline}
  \Bigl( \frac{4}{3}\,\rho_{{0}}\pi \,{r_{{0}}}^{3}-4\,{\frac {b   (
   ( r   ( t   )    ) ^{2}   ( \cos   ( \theta
   )    ) ^{2}+2\,r   ( t   ) b\cos   ( \theta
   ) +2\,{b}^{2}   ) \rho_{{0}}\pi }{   ( \cos   (
\theta   )    ) ^{3}}{{\rm e}^{-{\frac {r   ( t   )
\cos   ( \theta   ) }{b}}}}}
\\
+4\,{\frac {b   ( {r_{{0}}}^{2}
   ( \cos   ( \theta   )    ) ^{2}+2\,r_{{0}}b\cos
   ( \theta   ) +2\,{b}^{2}   ) \rho_{{0}}\pi }{   (
\cos   ( \theta   )    ) ^{3}}{{\rm e}^{-{\frac {r_{{0}}
\cos   ( \theta   ) }{b}}}}}  \Bigr  ) {\frac {\rm d}{{\rm d}t}}r
   ( t   )\\
    -\frac{4}{3}\,\rho_{{0}}\pi \,{r_{{0}}}^{3}v_{{0}}=0
\quad .
\label{eqndiffexp}
\end{multline}
An analytical solution does not exist
and we present the following
series solution of order 4 around $t_0$
\begin{multline}
r(t) =
r_{{0}}+ \left( t-{\it t_0} \right) v_{{0}}-3/2\,{\frac {{v_{{0}}}^{2}
 \left( t-{\it t_0} \right) ^{2}}{r_{{0}}}{{\rm e}^{-{\frac {r_{{0}}
\cos \left( \theta \right) }{b}}}}}
\\
+\frac{1}{2}\,{\frac {{v_{{0}}}^{3} \left(
t-{\it t_0} \right) ^{3}}{{r_{{0}}}^{2}b}{{\rm e}^{-{\frac {r_{{0}}\cos
 \left( \theta \right) }{b}}}} \left( 9\,{{\rm e}^{-{\frac {r_{{0}}
\cos \left( \theta \right) }{b}}}}b+r_{{0}}\cos \left( \theta \right)
-2\,b \right) } 
\\
+ \orderof \left ( t-t_0 \right)^4
\label{rtseriesexp}
\quad .
\end{multline}
A second approximate solution can be found by
deriving the velocity
from (\ref{eqndiffexp}):
\begin{equation}
v(r;r_0,v_0,\theta,b) =\frac{VN}{VD}
\end{equation}
where
\begin{equation}
VN= -{r_{{0}}}^{3}v_{{0}} \left( \cos \left( \theta \right)  \right) ^{3}
\quad ,
\end{equation}
and
\begin{multline}
VD=
3\,{{\rm e}^{-{\frac {r\cos   ( \theta   ) }{b}}}}   ( \cos
   ( \theta   )    ) ^{2}b{r}^{2}-3\,{{\rm e}^{-{\frac {r_{
{0}}\cos   ( \theta   ) }{b}}}}   ( \cos   ( \theta
   )    ) ^{2}{r_{{0}}}^{2}b-   ( \cos   ( \theta
   )    ) ^{3}{r_{{0}}}^{3}
   \\
   +6\,{{\rm e}^{-{\frac {r\cos
   ( \theta   ) }{b}}}}\cos   ( \theta   ) {b}^{2}r
   -6\,{
{\rm e}^{-{\frac {r_{{0}}\cos   ( \theta   ) }{b}}}}\cos
   ( \theta   ) r_{{0}}{b}^{2}
  \\
    +6\,{{\rm e}^{-{\frac {r\cos
   ( \theta   ) }{b}}}}{b}^{3}-6\,{{\rm e}^{-{\frac {r_{{0}}
\cos   ( \theta   ) }{b}}}}{b}^{3}
\quad .
\end{multline}
Given a function $f(r)$, the Pad\'e  approximant,
after \cite{Pade1892},
is
\begin{equation}
f(r)=\frac{a_{0}+a_{1}r+\dots+a_{p}r^{p}}{b_{0}+b_{1}%
r+\dots+b_{q}r^{q}}
\quad ,
\end{equation}
where the notation is the same of \cite{NIST2010}.
The coefficients $a_i$ and $b_i$
are found through Wynn's cross rule,
see \cite{Baker1975,Baker1996}
and our choice is $p=2$ and $q=1$.
The choice of  $p$ and $q$ is a compromise between
precision, high values for  $p$ and $q$, and
simplicity of the expressions to manage,
low values for  $p$ and $q$.
The inverse of the velocity expressed by the
the Pad\`e approximant
is
\begin{equation}
(\frac{1}{v(r)})_{2,1} = \frac{N21}{D21}
\quad .
 \end{equation}
 where 
\begin{multline}
N21=   ( r-r_{{0}}   )    ( 9\,{{\rm e}^{-{\frac {r_{{0}}\cos
   ( \theta   ) }{b}}}}br-9\,{{\rm e}^{-{\frac {r_{{0}}\cos
   ( \theta   ) }{b}}}}br_{{0}}
   \\+2\,\cos   ( \theta   ) r
r_{{0}}-2\,\cos   ( \theta   ) {r_{{0}}}^{2}-4\,br+10\,r_{{0}}b
   )
\end{multline}
and
\begin{equation}
D21=
2\,v_{{0}} \left( \cos \left( \theta \right) rr_{{0}}-\cos \left(
\theta \right) {r_{{0}}}^{2}-2\,br+5\,r_{{0}}b \right)
\end{equation}
 
 The above result allows deducing a solution $r_{2,1}$
expressed through the Pad\`e approximant
\begin{equation}
r(t)_{2,1} =
\frac
{
B+\sqrt {A}
}
{
9\,{{\rm e}^{-{\frac {r_{{0}}\cos \left( \theta \right) }{b}}}}b+2\,r_
{{0}}\cos \left( \theta \right) -4\,b
}
\label{rtpade}
\end{equation}
where
\begin{multline}
A=
 \left( \cos \left( \theta \right)  \right) ^{2}{r_{{0}}}^{2}{t}^{2}{v
_{{0}}}^{2}-2\, \left( \cos \left( \theta \right)  \right) ^{2}{r_{{0}
}}^{2}tt_{{0}}{v_{{0}}}^{2}+ \left( \cos \left( \theta \right)
 \right) ^{2}{r_{{0}}}^{2}{t_{{0}}}^{2}{v_{{0}}}^{2}
\\
-4\,\cos \left(
\theta \right) r_{{0}}b{t}^{2}{v_{{0}}}^{2}+8\,\cos \left( \theta
 \right) r_{{0}}btt_{{0}}{v_{{0}}}^{2}
-4\,\cos \left( \theta \right) r
_{{0}}b{t_{{0}}}^{2}{v_{{0}}}^{2}
\\
+54\,{{\rm e}^{-{\frac {r_{{0}}\cos
 \left( \theta \right) }{b}}}}r_{{0}}{b}^{2}tv_{{0}}-54\,{{\rm e}^{-{
\frac {r_{{0}}\cos \left( \theta \right) }{b}}}}r_{{0}}{b}^{2}t_{{0}}v
_{{0}}+6\,\cos \left( \theta \right) {r_{{0}}}^{2}btv_{{0}}
\\
-6\,\cos
 \left( \theta \right) {r_{{0}}}^{2}bt_{{0}}v_{{0}}+4\,{b}^{2}{t}^{2}{
v_{{0}}}^{2}-8\,{b}^{2}tt_{{0}}{v_{{0}}}^{2}
+4\,{b}^{2}{t_{{0}}}^{2}{v
_{{0}}}^{2}-12\,r_{{0}}{b}^{2}tv_{{0}}
\\
+12\,r_{{0}}{b}^{2}t_{{0}}v_{{0}
}+9\,{b}^{2}{r_{{0}}}^{2}
\end{multline}
and
\begin{multline}
B=
r_{{0}}tv_{{0}}\cos \left( \theta \right) -r_{{0}}t_{{0}}v_{{0}}\cos
 \left( \theta \right) +9\,{{\rm e}^{-{\frac {r_{{0}}\cos \left(
\theta \right) }{b}}}}br_{{0}}+2\,\cos \left( \theta \right) {r_{{0}}}
^{2}
\\
-2\,btv_{{0}}+2\,bt_{{0}}v_{{0}}-7\,r_{{0}}b
\quad .
\end{multline}

\begin{figure*}
\begin{center}
\includegraphics[width=7cm]{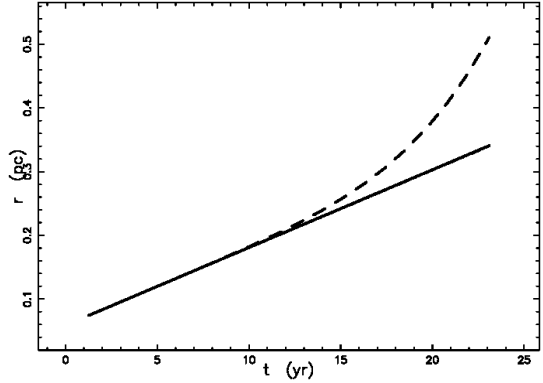}
\end {center}
\caption
{
Numerical   solution         (full   line) ,
power series solution        (dashed line)
and
Pad\'e  approximant solution (dot-dash-dot-dash line),
which is nearly equal to the numerical solution.
The parameters are
$r_0=0.06$ pc,
$t=21.86$\ yr,
$t_0 =0.1$\ yr,
$b= 0.011$ pc,
$\theta=0$,
and
$v_0\,=12000 \ km s^{-1}$.
}
\label{pade_exp}
    \end{figure*}

\begin{figure*}
\begin{center}
\includegraphics[width=7cm]{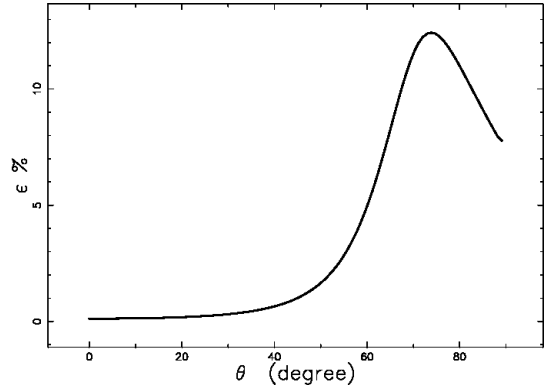}
\end {center}
\caption
{
Percentage error
of the Pad\'e  approximant solution compared
to the numerical solution, as a function of the
angle $\theta$ and
other  parameters as in Figure \ref{pade_exp}.
}
\label{pade_effi_exp}
    \end{figure*}

\begin{figure*}
\begin{center}
\includegraphics[width=7cm]{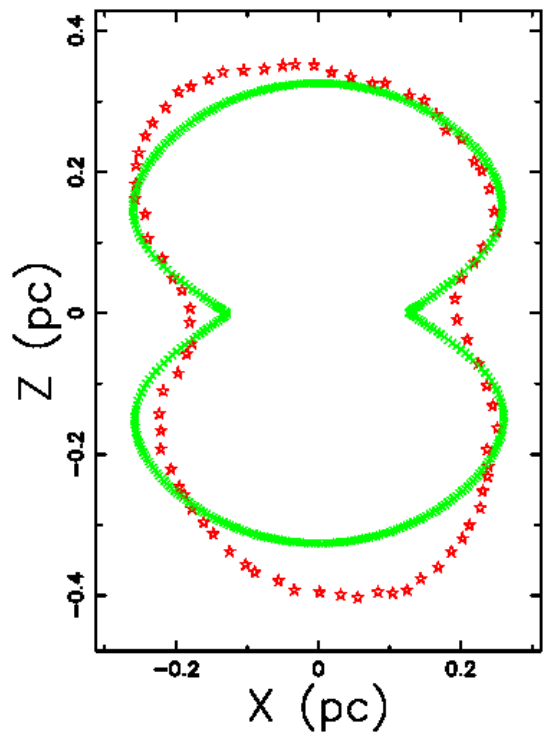}
\end {center}
\caption
{
Section of \sn1987a
in the $x-z$ plane with an exponential profile
(green points)
and observed profile
(red stars).
The parameters are the same  as Figure \ref{pade_exp}
and
$\epsilon_{\mathrm {obs}}=90.66\%$.
}
\label{cut_exp_1987a}
    \end{figure*}

Figure \ref{pade_exp} compares
the numerical solution,
the approximate series solution,
and the  Pad\'e  approximant solution.
The above figure clearly shows the limited range of validity
of the power series  solution.
The good agreement between  the Pad\'e  approximant solution
and numerical solution, in
Figure \ref{pade_exp}
the two solutions  can  not
be distinguished,
has a percentage error
\begin{equation}
\epsilon = \frac{\big | r(t) - r(t)_{2,1} \big |}
{r(t) } \times 100
\quad ,
\end{equation}
where $r(t)      $ is the numerical solution
and   $r(t)_{2,1}$ is the Pad\'e  approximant solution.
Figure \ref{pade_effi_exp} shows
the percentage error as a function of the
polar angle $\theta$.
Figure \ref{cut_exp_1987a} shows a cut  of  \sn1987a
in the $x-z$ plane evaluated with the numerical solution.
\subsection{Motion with a Gaussian profile}

\begin{figure*}
\begin{center}
\includegraphics[width=7cm]{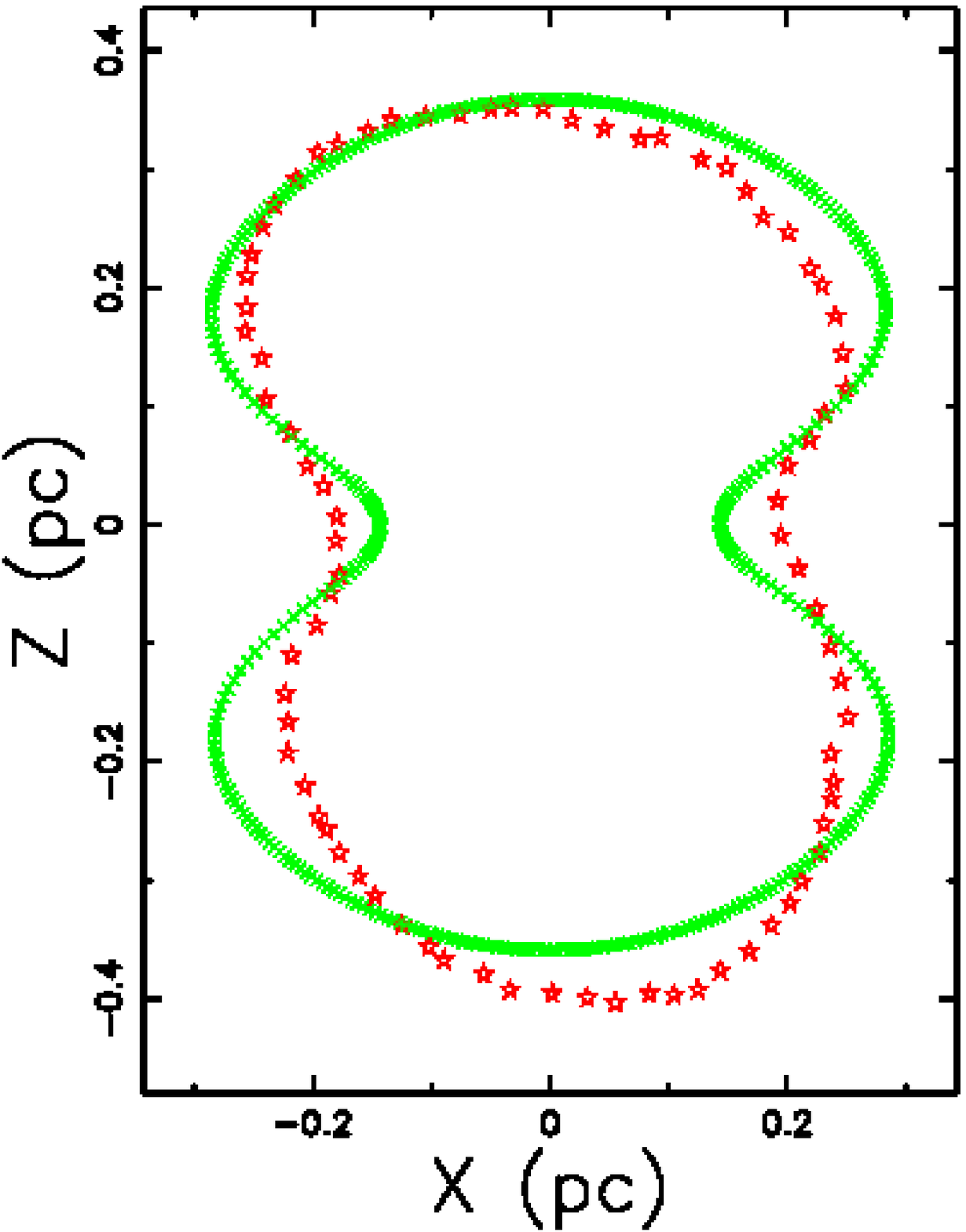}
\end {center}
\caption
{
Section of \sn1987a
in the $x-z$ plane with a Gaussian profile
(green points)
and observed profile
(red stars).
The parameters
$r_0=0.07$ pc,
$t=21.86$\ yr,
$t_0 =0.1$\ yr,
$b_0= 0.018$ pc,
$v_0=13000$\ km s$^{-1}$
give
$\epsilon_{\mathrm {obs}}=90.95\%$.
}
\label{cut_gauss_1987a}
    \end{figure*}

In the case of  a Gaussian density  profile
for the CSM
as given by Eq. (\ref{profgaussian}),
the differential equation
which models momentum conservation
is
\begin{multline}
   \bigl (\frac{4}{3}\,\rho_{{0}}\pi \,{r_{{0}}}^{3}+4\,\rho_{{0}}\pi \,    (
-{\frac {r    ( t    ) {b}^{2}}{    ( \cos    ( \theta
    )     ) ^{2}}{{\rm e}^{-\frac{1}{2}\,{\frac {    ( r    ( t
    )     ) ^{2}    ( \cos    ( \theta    )     ) ^{2}
}{{b}^{2}}}}}}
\\
+\frac{1}{2}\,{\frac {{b}^{3}\sqrt {\pi }\sqrt {2}}{    ( \cos
    ( \theta    )     ) ^{3}}{\rm erf}    (\frac{1}{2}\,{\frac {
\sqrt {2}\cos    ( \theta    ) r    ( t    ) }{b}}   )}
    )
  -4\,\rho_{{0}}\pi \,    ( -{\frac {r_{{0}}{b}^{2}}{    (
\cos    ( \theta    )     ) ^{2}}{{\rm e}^{-\frac{1}{2}\,{\frac {{r_{{0
}}}^{2}    ( \cos    ( \theta    )     ) ^{2}}{{b}^{2}}}}}}
\\
+
\frac{1}{2}\,{\frac {{b}^{3}\sqrt {\pi }\sqrt {2}}{    ( \cos    ( \theta
    )     ) ^{3}}{\rm erf}    (\frac{1}{2}\,{\frac {\sqrt {2}\cos
    ( \theta    ) r_{{0}}}{b}}   )}    )    \bigr  )
    {\frac
{\rm d}{{\rm d}t}}r    ( t    )
\\
- \frac{4}{3}\,\rho_{{0}}\pi \,{r_{{0}}}^{
3}v_{{0}}=0
\quad  .
\end{multline}
An analytical solution does not exist and
Figure \ref{cut_gauss_1987a}
shows a cut  of  \sn1987a
in the $x-z$ plane evaluated with a numerical solution.
\section{The relativistic thin layer approximation}
\label{secrelativistic}

The conservation of relativistic momentum
in spherical coordinates
along  the  solid angle  $\Delta \Omega$
in the framework of the thin
layer approximation  gives
\begin{equation}
{\frac {M \left( t ;\theta\right) \beta}{\sqrt {1-{\beta}^{2}}}}
\Delta \Omega
={\frac {M_{
{0}} \left( t_{{0}} \right) \beta0}{\sqrt {1-{\beta0}^{2}}}}
\Delta \Omega
\quad .
\end{equation}
where $M_0(r_0)$ and $M(r)$ are the swept
masses at $r_0$ and $r$,
$\beta=\frac{v}{c}$ , $\beta_0=\frac{v_0}{c}$,
and $v_0$ and $v$ are the velocities of the thin
layer at $r_0$ and $r$.
We have chosen as units pc for distances and
yr as time and therefore   the speed of light is
$c=0.306$\ pc\ yr$^{-1}$.

\subsection{Relativistic motion with a hyperbolic profile}

In the case of a hyperbolic density profile for the CSM
as given by Eq. (\ref{profhyperbolic}),
the differential equation
which models the momentum conservation
is
\begin{equation}
 \left(\frac{4}{3}\,\rho_{{0}}\pi \,{r_{{0}}}^{3}+2\,{\frac {\rho_{{0}}{\it
z_0}\,\pi \, \left( -{r_{{0}}}^{2}+ \left( r \left( t \right)  \right)
^{2} \right) }{\cos \left( \theta \right) }} \right) {\frac {\rm d}{
{\rm d}t}}r \left( t \right) -\frac{4}{3} \,\rho_{{0}}\pi \,{r_{{0}}}^{3}v_{{0}
}=0
\quad ,
\end{equation}
where the initial  conditions
are  $r=r_0$  and   $v=v_0$
when $t=t_0$.

\subsection{Relativistic motion with a hyperbolic profile}

\begin{figure*}
\begin{center}
\includegraphics[width=7cm]{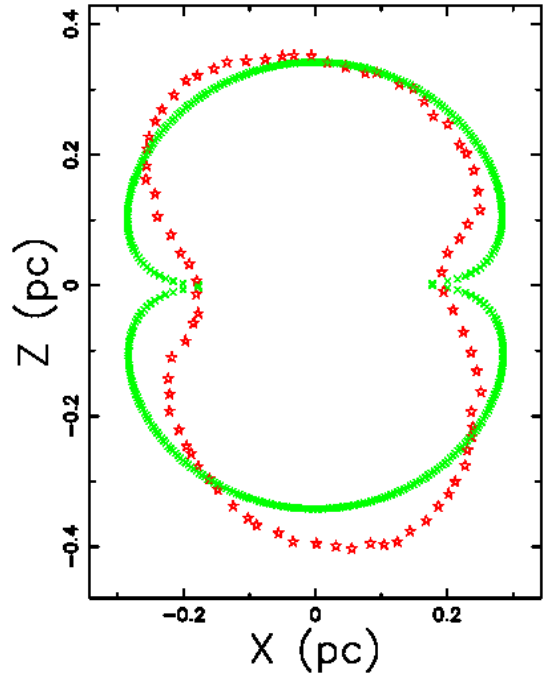}
\end {center}
\caption
{
Section of \sn1987a
in the $x-z$ plane with a hyperbolic profile: relativistic case
(green points)
and observed profile (red stars).
The parameters
$r_0=0.1$ pc,
$z_0=0.003$\ pc,
$t=21.86$\ yr,
$t_0 =0.1$ yr,
$v_0\,=78000$\ km s$^{-1}$,
and $\beta_0=0.26$
give
$\epsilon_{\mathrm {obs}}=88.68\%$.
}
\label{cut_hyper_rel_1987a}
    \end{figure*}

In the case of a hyperbolic density profile for the CSM
as given by Eq. (\ref{profhyperbolic}),
the differential equation
which models the relativistic momentum conservation
is
\begin{equation}
\frac
{
2\,\rho_{{0}}\pi \, \left( 2\,{r_{{0}}}^{3}\cos \left( \theta \right)
+3\,{\it z0}\, \left( r \left( t \right)  \right) ^{2}-3\,{\it z0}\,{r
_{{0}}}^{2} \right) {\frac {\rm d}{{\rm d}t}}r \left( t \right)
}
{
3\,\cos \left( \theta \right) c\sqrt {1-{\frac { \left( {\frac {\rm d}{
{\rm d}t}}r \left( t \right)  \right) ^{2}}{{c}^{2}}}}
}
-
\frac
{
4\,\rho_{{0}}\pi \,{r_{{0}}}^{3}\beta_{{0}}
}
{
3\,\sqrt {1-{\beta_{{0}}}^{2}}
}
=0
\label{eqndiffrelhyper}
\quad  .
\end{equation}
The velocity expressed in terms of $\beta$ can be derived
from the above equation:
\begin{equation}
\beta = \frac
{
2\,{r_{{0}}}^{3}\cos \left( \theta \right) \beta_{{0}}
}
{
D
}
\quad ,
\end{equation}
where
\begin{multline}
D=
\Bigl (
12\,{r_{{0}}}^{5}{\beta_{{0}}}^{2}{\it z0}\,\cos \left( \theta
 \right) -12\,{r_{{0}}}^{3}{\beta_{{0}}}^{2}{r}^{2}{\it z0}\,\cos
 \left( \theta \right) 
\\
+4\,{r_{{0}}}^{6} \left( \cos \left( \theta
 \right)  \right) ^{2}-9\,{r_{{0}}}^{4}{\beta_{{0}}}^{2}{{\it z0}}^{2} \\
+18\,{r_{{0}}}^{2}{\beta_{{0}}}^{2}{r}^{2}{{\it z0}}^{2}-9\,{\beta_{{0
}}}^{2}{r}^{4}{{\it z0}}^{2}-12\,{r_{{0}}}^{5}{\it z0}\,\cos \left(
\theta \right) 
\\
+12\,{r_{{0}}}^{3}{r}^{2}{\it z0}\,\cos \left( \theta
 \right)
   +9\,{r_{{0}}}^{4}{{\it z0}}^{2}-18\,{r_{{0}}}^{2}{r}^{2}{{
\it z0}}^{2}+9\,{r}^{4}{{\it z0}}^{2} \Bigr )^{\frac{1}{2}}
\quad .
\end{multline}
An analytical solution
of (\ref{eqndiffrelhyper})
does not exist, but we present the following
series solution of order three around $t_0$:
\begin{equation}
r(t)=r_{{0}}+\beta_{{0}}c \left( t-{\it t0} \right) +\frac{3}{2}  {\frac {{c}^{2}
 \left( {\beta_{{0}}}^{2}-1 \right) z_{{0}}{\beta_{{0}}}^{2} \left( t-
{\it t0} \right) ^{2}}{{r_{{0}}}^{2}\cos \left( \theta \right) }}
+ \orderof \left ( t-t_0 \right)^3
\label{rtserieshyperrel}
\quad .
\end{equation}
Figure \ref{cut_hyper_rel_1987a} shows the numerical solution
for  \sn1987a
in the $x-z$ plane.

\subsection{Relativistic motion with an exponential profile}

In the case of an exponential density profile for the CSM
as given by Eq. (\ref{profexponential}),
the differential equation
which models the relativistic momentum conservation
is
\begin{equation}
\frac{N\,{\frac {\rm d}{{\rm d}t}}r   ( t
   )}
{
3\, \left( \cos \left( \theta \right)  \right) ^{3}c\sqrt {1-{\frac {
 \left( {\frac {\rm d}{{\rm d}t}}r \left( t \right)  \right) ^{2}}{{c}
^{2}}}}
}
-
\frac
{
4\,\rho_{{0}}\pi \,{r_{{0}}}^{3}\beta_{{0}}
}
{
3\,\sqrt {1-{\beta_{{0}}}^{2}}
}
=0
\quad  ,
\end{equation}
where
\begin{multline}
N=-4\,\rho_{{0}}\pi \,   \Bigl ( 3\,   ( r   ( t   )    ) ^{2
}   ( \cos   ( \theta   )    ) ^{2}{{\rm e}^{-{\frac {r
   ( t   ) \cos   ( \theta   ) }{b}}}}b-   ( \cos
   ( \theta   )    ) ^{3}{r_{{0}}}^{3}
   \\
   -3\,   ( \cos
   ( \theta   )    ) ^{2}{{\rm e}^{-{\frac {r_{{0}}\cos
   ( \theta   ) }{b}}}}{r_{{0}}}^{2}b
     +6\,r   ( t   )
\cos   ( \theta   ) {{\rm e}^{-{\frac {r   ( t   ) \cos
   ( \theta   ) }{b}}}}{b}^{2}
   \\
   -6\,\cos   ( \theta   ) {
{\rm e}^{-{\frac {r_{{0}}\cos   ( \theta   ) }{b}}}}r_{{0}}{b}^
{2}
-6\,{{\rm e}^{-{\frac {r_{{0}}\cos   ( \theta   ) }{b}}}}{b}
^{3}+6\,{{\rm e}^{-{\frac {r   ( t   ) \cos   ( \theta
   ) }{b}}}}{b}^{3}  \Bigr )
   \label{eqndiffrelexp}
\quad  .
\end{multline}
An analytical solution
of (\ref{eqndiffrelexp})
does not exist, so we present the following
series solution of order three around $t_0$:
\begin{multline}
r(t)=
\\
r_{{0}}+c\beta_{{0}} \left( t-{\it t0} \right) +\frac{3}{2}\,{\frac {{\beta_{{0
}}}^{2}{c}^{2} \left( {\beta_{{0}}}^{2}-1 \right)  \left( t-{\it t0}
 \right) ^{2}}{r_{{0}}}{{\rm e}^{-{\frac {r_{{0}}\cos \left( \theta
 \right) }{b}}}}}
+ \orderof \left ( t-t_0 \right)^3
\label{rtseriesexprrel}
\quad  .
\end{multline}

\begin{figure*}
\begin{center}
\includegraphics[width=7cm]{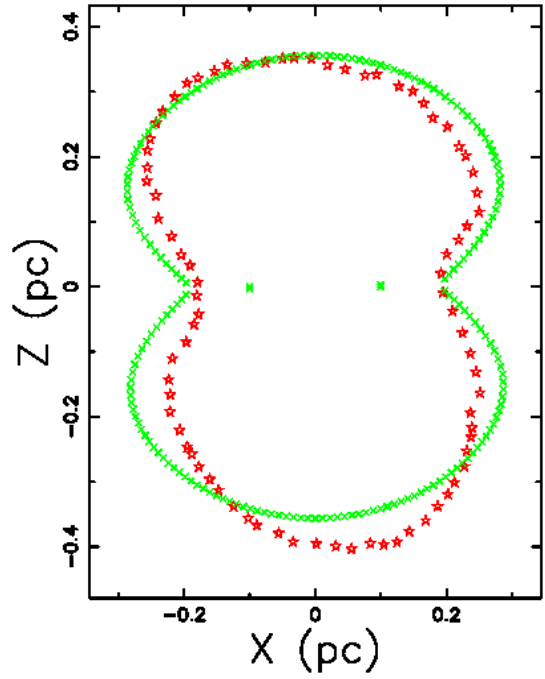}
\end {center}
\caption
{
Section of \sn1987a
in the $x-z$ plane with an exponential
profile: relativistic case
(green points)
and observed profile (red stars).
The parameters
$r_0$=0.1  pc,
$b=0.02$\ pc,
$t=21.86$\ yr,
$t_0 =0.1$~yr,
$v_0\,=70000$\ km s$^{-1}$,
and $\beta_0=0.233$
give
$\epsilon_{\mathrm {obs}}=90.23\%$.
}
\label{cut_exp_rel_1987a}
    \end{figure*}

Figure \ref{cut_exp_rel_1987a} shows the numerical solution
for  \sn1987a
in the $x-z$ plane.

\section{Conclusions}

{\bf Type of medium:}

We have selected four density profiles, which decrease
with the distance ($z$-axis) from the equatorial plane.
The integral which evaluates the swept mass increases
in complexity according to the following
sequence of density profiles: hyperbolic,
power law,
exponential, and
Gaussian.

{\bf Classical thin layer}

The application of the thin layer approximation
with different profiles produces differential equations
of the first order.
The solution of the first order differential equation
can be  analytical in the classical case characterized by a
hyperbolic density profile,
see (\ref{rtanalyticalhyper}),
and numerical  in all other cases.
We also evaluated the approximation of the solution
as a power law series,
see
(\ref{rtseriesexp}),
or using the  Pade approximant, see (\ref{rtpade}):
the differences between the two approximations
are outlined in Figure \ref{pade_exp}.

{\bf Relativistic thin layer}

The application of the thin layer approximation
to the relativistic case produces first order differential
equations which can be solved only numerically
or as a power series,
see
(\ref{rtseriesexprrel})
and
(\ref{rtserieshyperrel}).

{\bf The astrophysical case}

The application of the theory here developed
is connected with a clear definition of the
advancing SN's surface in 3D.
We have concentrated the analysis on \sn1987a
with the cuts of the advancing surface
in the $x-z$ plane when the $z$ axis is in front of the observer.
Another choice of the
point of view of the observer would complicate
the situation, and a comparison between theory and
observation then requires the introduction of
the three
Euler angles which characterize
the observer,
see the rotated advancing surface of \sn1987a
shown in Figure \ref{3dsurfacehyper}.
Is interesting note that
the rotation of the observed image
with  the polar axis aligned with the $z$-direction
has been done for the
Homunculus
nebula,
see  Figure 4 in \cite{Smith2006},
but only an approximate section
of the H-$\alpha$ imaging  connected with
\sn1987a has been already reported,
see Figure 5
in  \cite{France2015}.
A more precise definition of the section of \sn1987a
will help the theoretical determination of the parameters
maximizing the observational reliability,
$\epsilon_{\mathrm {obs}}$,
see \ref{efficiencymany}.
Is important note that the observational reliability
gives already an acceptable result and
$\epsilon_{\mathrm {obs}}$
lies within the interval [88\%--92\%] for the six models, four classical and two relativistic,
here analyzed.

\end{document}